\documentstyle[12pt]{article}
\begin{document}
\hsize = 5.7 in
\vsize =11.7 in
\hoffset=0.1 in
\voffset=-0.5 in
\baselineskip=20pt
\newcommand{\be}{\begin{equation}}
\newcommand{\ee}{\end{equation}}
\newcommand{\bea}{\begin{eqnarray}}
\newcommand{\eea}{\end{eqnarray}}
\newcommand{\bfx}{\mbox{\boldmath $x$}}
\newcommand{\bfy}{\mbox{\boldmath $y$}}
\newcommand{\bfz}{\mbox{\boldmath $z$}}
\newcommand{\bfw}{\mbox{\boldmath $w$}}
\newcommand{\bfR}{\mbox{\boldmath $R$}}
\newcommand{\bfC}{\mbox{\boldmath $C$}}
\newcommand{\bfE}{\mbox{\boldmath $E$}}
\newcommand{\bfEG}{{\mbox{\boldmath $E$}}^{(G)}}
\newcommand{\bfEbar}{\overline{\mbox{\boldmath $E$}}}
\newcommand{\bfe}{\mbox{\boldmath $e$}}
\newcommand{\bfeg}{{\mbox{\boldmath $e$}}^{(g)}}
\newcommand{\ZERO}{\mbox{\boldmath $0$}}
\newcommand{\Pk}{{{\cal P}^{k}}}
\newcommand{\Pl}{{{\cal P}^{l}}}
\newcommand{\pk}{{{\scriptstyle{{\cal P}}}^{k}}}
\newcommand{\spk}{{{\scriptscriptstyle{{\cal P}}}^{k}}}
\newcommand{\VPk}{{}^{\Pk}V}
\newcommand{\VPl}{{}^{\Pl}V}
\newcommand{\FPk}{{}^{\Pk}F}
\newcommand{\FPl}{{}^{\Pl}F}
\newcommand{\VnotPk}{{\VPk}_{0}}
\newcommand{\VnotPl}{{\VPl}_{0}}
\newcommand{\vpk}{{}^{\spk}v}
\newcommand{\Jk}{{{\cal J}^{k}}}
\newcommand{\Jm}{{{\cal J}^{m}}}
\newcommand{\jk}{{{\scriptstyle{{\cal J}}}^{k}}}
\newcommand{\sjk}{{{\scriptscriptstyle{{\cal J}}}^{k}}}
\newcommand{\VJk}{{}^{\Jk}V}
\newcommand{\VJm}{{}^{\Jm}V}
\newcommand{\VnotJk}{{\VJk}_{0}}
\newcommand{\vjk}{{}^{\sjk}v}
\newcommand{\Kk}{{{\cal K}^{k}}}
\newcommand{\Kl}{{{\cal K}^{l}}}
\newcommand{\kk}{{{\scriptstyle{{\cal K}}}^{k}}}
\newcommand{\skk}{{{\scriptscriptstyle{{\cal K}}}^{k}}}
\newcommand{\skm}{{{\scriptscriptstyle{{\cal K}}}^{m}}}
\newcommand{\skn}{{{\scriptscriptstyle{{\cal K}}}^{n}}}
\newcommand{\KKk}{{}^{\Kk}K}
\newcommand{\KKl}{{}^{\Kl}K}
\newcommand{\kkk}{{}^{\skk}k}
\newcommand{\kkm}{{}^{\skm}k}
\newcommand{\kkn}{{}^{\skn}k}
\newcommand{\Pnot}{{{\cal P}^{0}}}
\newcommand{\pnot}{{{\scriptstyle{{\cal P}}}^{0}}}
\newcommand{\ZKk}{{}^{\Kk}Z}
\newcommand{\ZKl}{{}^{\Kl}Z}
\newcommand{\zkk}{{}^{\skk}z}
\newcommand{\Abar}{\overline{A}}
\newcommand{\Bbar}{\overline{B}}
\newcommand{\Cbar}{\overline{C}}
\newcommand{\Dbar}{\overline{D}}
\newcommand{\Ebar}{\overline{E}}
\newcommand{\del}{\tilde{\nabla}}
\newcommand{\Q}{{\cal Q}}
\newcommand{\Qs}{{\scriptstyle{\cal Q}}}
\newcommand{\C}{{\cal C}}
\newcommand{\Cs}{{\scriptstyle{\cal C}}}
\newcommand{\R}{{\cal R}}
\newcommand{\r}{{\tilde{\cal R}}}
\newcommand{\F}{{\cal F}}
\newcommand{\Fphys}{{\cal F}_{\rm phys}}
\newcommand{\Psiphys}{\Psi_{\rm phys}}
\newcommand{\lng}{\ln\sqrt{\gamma}}
\newcommand{\lb}{\tilde{\Delta}}
\newcommand{\G}{{\cal G}}
\newcommand{\Gbar}{\overline{\cal G}}
\newcommand{\Obar}{\overline{\Omega}}
\newcommand{\D}{\tilde{D}}
\newcommand{\tr}{{\rm tr}}

A page for thought.
\clearpage
\pagenumbering{arabic}
\begin{center}
{\large\bf Dirac versus Reduced Quantization of the Poincar\'{e}
Symmetry in Scalar Electrodynamics}\\
\vspace{15 pt}
R.J. Epp\\
{\footnotesize\it Physics Department, University of Winnipeg,
Winnipeg, Manitoba R3B 2E9 Canada.}\\
\vspace{10 pt}
G. Kunstatter\\
{\footnotesize\it Physics Department, University of Manitoba,
Winnipeg, Manitoba R3T 2N2 Canada.}\\
{\footnotesize\it Winnipeg Institute for Theoretical Physics,
Physics Department, University of Winnipeg, Winnipeg, Manitoba R3B
2E9 Canada.}\\
\end{center}
\vspace{15 pt}
{\noindent
{\large\bf Abstract} :
The generators of the Poincar\'{e}
symmetry of scalar electrodynamics are quantized in the functional
Schr\"{o}dinger representation.  We show that the factor ordering
which corresponds to (minimal) Dirac
quantization preserves the Poincar\'{e} algebra, but (minimal)
reduced quantization does not.  In the latter, there is a van Hove
anomaly in the boost-boost commutator, which we evaluate explicitly
to lowest order in a heat kernel expansion using zeta function
regularization. We illuminate the crucial role played by the gauge
orbit volume element in the analysis. Our results demonstrate that
preservation of extra symmetries at the quantum level is sometimes
a useful criterion to select
between inequivalent, but nevertheless self-consistent,
quantization schemes.}
\vspace{1 cm}\\
WIN-94-03

\clearpage
\section{Introduction}

\par
There has been a long standing debate in the literature
concerning the Dirac versus reduced quantization of gauge theories.
In Dirac quantization, one constructs quantum operators on the full
space of fields prior to reducing to the physical degrees of
freedom. The gauge constraints are then realized as operator
constraints on physical states. Reduced quantization, on the other
hand, as the name suggests, constructs quantum operators for
physical observables only. Dirac quantization is simpler in the
sense that the full space of fields is usually endowed
with a flat configuration space metric. It has the disadvantage,
however, of including supposedly unphysical information into the
quantization scheme. It is well known that these two approaches to
quantization generally lead to distinct quantum systems
\cite{Ashtekar,Kuchar1,Kuchar2,Romano,Schleich,Kunstatter1}, and
that the difference can be understood as a factor ordering
ambiguity involving the volume element on the gauge orbits
\cite{Kunstatter1}.  Now it can happen that both approaches are
self consistent, and so even though the respective
Hamiltonians may yield different spectra, there is no {\em
internal} criterion with which
to select the correct factor ordering. This has been illustrated
by
Kucha\v{r} \cite{Kuchar2} using a finite dimensional model, which
we will refer to as the helix model.
\par
The purpose of the present paper is to examine in detail the
quantization of a field theoretic version of Kucha\v{r}'s helix
model, namely scalar electrodynamics in flat
spacetime.  An important distinction between the two
models in the present context is that scalar electrodynamics
contains a symmetry not present in
the helix model: Poincar\'{e} invariance.  Our principle
contribution is to show that Poincar\'{e} invariance at the quantum
level is sensitive to this factor ordering ambiguity, and provides
a suitable internal criterion:  minimal\footnote{This term is
defined later.} Dirac quantization passes, whereas minimal reduced
quantization appears to fail.
\par
The paper is organized as follows.  In section 2 we review both the
Lagrangian and Hamiltonian analyses of scalar electrodynamics,
chiefly to introduce notation.  This is followed in the next
section by a discussion of the classical Poincar\'{e} symmetry: we
write down the classical Poincar\'{e} charges on the full phase
space, and verify that these generate the Poincar\'{e} algebra, up
to `off-shell' pieces which vanish on the constraint surface.  Then
in section 4 we quantize the Poincar\'{e} generators in the
functional Schrodinger representation and show that minimal Dirac
quantization preserves the Poincar\'{e} symmetry when acting on the
physical state space.  This involves showing that all potential van
Hove anomalies \cite{vanHove,Groenewold,Abraham} vanish, and that
the off-shell pieces mentioned above annihilate physical states
when quantized.
In section 5 we turn to minimal reduced quantization, and
demonstrate that it does {\em not} preserve the Poincar\'{e}
symmetry: there exists a van Hove anomaly in, for example, the
boost-boost commutator.  This calculation involves zeta function
regularization (via heat kernel techniques) \cite{Hawking,DeWitt}
of (the log of) the gauge orbit volume element.  Finally, in
section
6
we compare both Dirac and reduced quantizations (acting on the same
physical state space) to clarify why minimal Dirac quantization
succeeds, while minimal reduced does not.  It is clear that the
volume element on the gauge orbits plays a pivotal role.\\[5pt]
\section{Scalar Electrodynamics}
\par
The Lagrangian density for scalar
electrodynamics in globally Lorentzian spacetime is
\be
{\cal L}=
{1 \over 2}( D_\mu \varphi )\overline{( D^\mu \varphi )} -
U-{1 \over 4} F_{\mu \nu} F^{\mu \nu} ,
\label{eq: Lagrangian density}
\ee
where we use spacetime signature $(+---)$, and the indices
$\mu,\nu$ run from 0 to 3.  $\varphi:=\xi+i\eta$ is a complex
scalar field and $U$ is a potential,
for example a mass or self interaction term,
which depends only on $\vert \varphi\vert$.
The covariant derivative is $D_\mu := \partial_\mu + i e A_\mu$,
with corresponding electromagnetic field strength
$F_{\mu\nu} := \partial_\mu A_\nu -\partial_\nu A_\mu$.

\par
Fixing an inertial frame, and using DeWitt's condensed notation,
the Lagrangian $L(t):=\int d^3x {\cal L}(t,\bfx )$
can be cast into the form:
\be
L(\lambda,Q,\dot{Q})={1 \over 2}G_{AB}(Q)\left(\dot{Q}^{A}-\lambda^
{\alpha}
\phi_{\alpha}^{A}(Q)\right)\left(\dot{Q}^B-\lambda^{\beta}
\phi_{\beta}^{B}(Q)\right)-V(Q) ,
\label{eq: Lagrangian}
\ee
In the above, the configuration of the system at time instant
$t$ is represented by a point in the configuration space, $M$, with
coordinates
\be
Q^{A}:=\left(A_{i}(\bfx ),\xi(\bfx ),\eta(\bfx )
\right),
\label{eq: coords on M}
\ee
where the index $A$ runs over discrete values (including the
spatial index $i=1,2,3$), as well as the continuum $\alpha:=\bfx\in
\bfR^3$.  Repeated indices imply summation and/or integration, as
appropriate.The overdot on the velocities $\dot{Q}^A$ indicates
time
derivative, but the time argument has been suppressed.  The
time-like part of the vector potential plays the role of a Lagrange
multiplier: $\lambda^{\alpha}:=-eA_{0}(\bfx )$.  The kinetic energy
term in the Lagrangian induces a flat metric on $M$, with
components
\be
G_{AB}(Q):=\left( \begin{array}{ccc}
                   \delta^{ij} & 0 & 0 \\
                    0 & 1 & 0 \\
                    0 & 0 & 1 \end{array}
                    \right)\delta(\bfx-\bfy)
\label{eq: metric}
\ee
in the Cartesian coordinates $Q^{A}$.  The potential term in $L$
is
\be
V(Q):=\int d^{3}x\,\left\{ {1\over 4}(F_{ij})^2
+{1\over 2}(\partial_i \xi - eA_i \eta)^2
+{1\over 2}(\partial_i \eta + eA_i \xi)^2
+U
\right\}.
\label{eq: potential}
\ee

\par
Gauge transformations on $M$ are generated by the `gauge vector
fields' $\phi_{\alpha}=\phi_{\alpha}^{A}\partial/\partial Q^{A}$,
whose
components in the Cartesian coordinates are
\be
\phi_{\beta}^{A}(Q)=
\left(-{{1}\over{e}}\partial_{x^i}\delta(\bfx-\bfy), -\eta(\bfx
)\delta(\bfx-\bfy),
\xi(\bfx )\delta(\bfx-\bfy) \right) .
\label{eq: gauge vectors}
\ee
The gauge vector fields are linearly independent (except on the
$\xi=\eta=0$ `axis'), and their Lie bracket algebra:
\be
[\phi_\alpha,\phi_\beta]=f^\gamma_{\alpha\beta}\phi_{\gamma}=0
\label{eq: gauge algebra}
\ee
is, of course, abelian.

\par
The phase space $\Gamma=T^{*}M$.  A straightforward Hamiltonian
analysis (see, e.g. \cite{Sundermeyer}) yields the canonical
Hamiltonian
\be
H(\lambda,Q,P)={1\over 2}G^{AB}(Q)P_{A}P_{B}+V(Q)+\lambda^{\alpha}
C_{\alpha}(Q,P),
\label{eq: Hamiltonian}
\ee
in which $G^{AB}$ denotes the matrix inverse of $G_{AB}$.  The
momenta $P_{A}$, conjugate to $Q^{A}$, are (in uncondensed
notation)
\bea
\Pi_{A_{i}(\bfx)}&=&{{\delta L}\over{\delta \dot{A}_{i}(\bfx)}}
=\dot{A}_{i}(\bfx)-\partial_{x^i}A_{0}(\bfx)=F_{0i}(\bfx),
\\
\Pi_{\xi(\bfx)}&=&{{\delta L}\over{\delta \dot{\xi}(\bfx)}}
=\dot{\xi}(\bfx)-eA_{0}(\bfx)\eta(\bfx),
\\
\Pi_{\eta(\bfx)}&=&{{\delta L}\over{\delta \dot{\eta}(\bfx)}}
=\dot{\eta}(\bfx)+eA_{0}(\bfx)\xi(\bfx),
\eea
where ${{\delta}/{\delta \dot{A}_{i}(\bfx)}}$, etc., denotes
functional derivative.

\par
The Lagrange multipliers $\lambda^{\alpha}$ enforce the Gauss law
constraints
\be
C_{\alpha}(Q,P) := \phi_{\alpha}^{B}(Q)P_B
={{1}\over{e}}\partial_{x^i}\Pi_{A_i (\bfx)}-\eta
(\bfx)
\Pi_{\xi (\bfx)}+\xi (\bfx) \Pi_{\eta (\bfx)}\approx 0,
\label{eq: constraints}
\ee
which defines a constraint surface $\Gamma_{C}\subset\Gamma$.  It
turns out that the $\phi_{\alpha}$ are Killing vectors and that the
potential $V$ is constant along the gauge orbits in $M$, conditions
which, together with
(\ref{eq: gauge algebra}), are sufficient to guarantee that
$\Gamma_{C}$ is invariant under time evolution.

\section{Classical Poincar\'{e} Symmetry}

\par
Integrating the Lagrangian density $\cal L$ in
(\ref{eq: Lagrangian density}) over spacetime gives the action,
whose functional derivative with respect to the spacetime metric
(evaluated at the globally Lorentzian metric, $\eta_{\mu\nu}$)
yields the symmetric and conserved energy-momentum tensor
\be
T^{\mu\nu}=
( D^{(\mu} \varphi )\overline{( D^{\nu)} \varphi )} -
F^{\rho\mu} F_{\rho}^{\nu}-\eta^{\mu\nu}{\cal L},
\ee
where parentheses on the indices denotes symmetrization.  Together
with the Poincar\'{e} group of isometries, this then leads to the
conserved Poincar\'{e} charges
\bea
{\cal P}^{\mu}&:=&\int
d^{3}x\,T^{0\mu},
\\
{\cal J}^{\mu\nu}&:=&\int
d^{3}x\,\left(x^{\mu}T^{0\nu}-x^{\nu}T^{0\mu}\right).
\eea
As we shall see shortly, as dynamical variables on $\Gamma$ these
constants of the motion canonically generate transformations on the
classical states which realize the Poincar\'{e} algebra, at least
when acting on the constraint surface $\Gamma_{C}$.

\par
We find, in terms of the phase space variables,
\bea
T^{00}&=&{1\over2}\left(\Pi_{A_{i}}^2 + \Pi_{\xi}^2 + \Pi_{\eta}^2
\right)+{\cal V}(A_{i},\xi,\eta),
\\
T^{0i}&=&-\Pi_{A_j}F_{ij}-\Pi_{\xi}\left(\partial_{i}\xi-
eA_{i}\eta\right)-
\Pi_{\eta}\left(\partial_{i}\eta+eA_{i}\xi\right),
\eea
where ${\cal V}(A_{i},\xi,\eta)$ is the integrand in
(\ref{eq: potential}).  Thus the generator of time translation
\be
{\cal P}^{0}=\C_{2}({1\over2}G^{-1})+\C_{0}(V)
\label{eq: notation for charges 1}
\ee
in condensed notation, where $\C_{s}(S):=S^{A_{1}\cdots
A_{s}}(Q)P_{A_1}\cdots P_{A_s}$ denotes the homogeneous classical
dynamical variable on $\Gamma$ associated with a symmetric
contravariant valence $s$ tensor field $S$ on $M$ (cf
(\ref{eq: Hamiltonian})).

\par
After an integration by parts the spatial translation generators
turn out to be
\bea
{\cal P}^{k}&=&\int d^{3}x\,\left\{-\Pi_{A_l}\partial_{k}A_{l}-
\Pi_{\xi}\partial_{k}\xi-
\Pi_{\eta}\partial_{k}\eta-eA_{k}C_{\alpha}\right\}
\nonumber
\\
&=:&\C_{1}({}^{k}X),
\eea
where we read off the vector field components
\be
{}^{k}X^{A}(Q)=\left(-\partial_{x^k}A_{i}(\bfx),-
\partial_{x^k}\xi(\bfx),-\partial_{x^k}\eta(\bfx)\right)
-e\int d^{3}z\, A_{k}(\bfz)\phi_{\gamma}^{A}(Q).
\ee
Here $\gamma:=\bfz$ is included in the integration.  Similarly, the
spatial rotation generators
\bea
{\cal J}^{k}&:=&{1\over2}[kmn]{\cal J}^{mn}
\nonumber
\\
&=&\int d^{3}x\,[kmn]\left\{x^{m}\left[-
\Pi_{A_i}\partial_{n}A_{i}-\Pi_{\xi}\partial_{n}\xi-
\Pi_{\eta}\partial_{n}\eta-eA_{n}C_{\alpha}\right]-
\Pi_{A_m}A_{n}\right\}
\nonumber
\\
&=:&\C_{1}({}^{k}Y),
\eea
where [kmn] is the completely antisymmetric symbol in three
dimensions, with [123]=1, and
\bea
{}^{k}Y^{A}(Q)&=&
[kmn]\left(
-x^{m}\partial_{x^n}A_{i}(\bfx)-\delta_{i}^{m}A_{n}(\bfx),
-x^{m}\partial_{x^n}\xi(\bfx),
-x^{m}\partial_{x^n}\eta(\bfx)\right)
\nonumber
\\
&&-e\int d^{3}z\,[kmn]z^{m}A_{n}(\bfz)\phi_{\gamma}^{A}(Q).
\eea

\par
Finally, the boost generators
\be
{\cal K}^{k}:={\cal J}^{0k}=
-\C_{2}({1\over2}{}^{k}K)-
\C_{0}({}^{k}V)+t{\cal P}^{k},
\label{eq: notation for charges 4}
\ee
where the boost tensors
\be
{}^{k}K^{AB}(Q)=
\left( \begin{array}{ccc}
                   \delta_{ij} & 0 & 0 \\
                    0 & 1 & 0 \\
                    0 & 0 & 1 \end{array}
                    \right){1\over2}(x^{k}+y^{k})\delta(\bfx-\bfy),
\label{eq: boost tensor}
\ee
and the boost potentials
\be
{}^{k}V(Q)=\int d^{3}x\,x^{k}{\cal V}(A_{i},\xi,\eta),
\label{eq: boost potentials}
\ee
which are analogous to the potential $V(Q)$ in the Hamiltonian.

\par
For future reference we record here some properties of the various
tensors associated with the Poincar\'{e} charges.  First, the Lie
derivative with respect to $\phi_\alpha$ of every valence zero,
one, and two tensor that occurs in ${\cal P}^{0}$, ${\cal P}^{k}$,
${\cal J}^{k}$, ${\cal K}^{k}$ vanishes.\footnote{For the explicit
calculation refer to \cite{thesis}.}  This is sufficient (but not
necessary) to guarantee that the Poincar\'{e} charges are classical
observables, that is, gauge invariant on the constraint surface
$\Gamma_{C}$.  Next, the boost tensors ${}^{k}K$ have field
independent components in the Cartesian coordinates (see
(\ref{eq: boost tensor})), and so are, in fact, covariantly
constant (cf (\ref{eq: metric})).\footnote{This implies they are
Killing, and in involution, which, modulo terms that vanish on
$\Gamma_{C}$, is necessary for the Poincar\'{e} algebra to close.}
Finally, the spatial translation and rotation vectors ${}^{k}X$ and
${}^{k}Y$, while not Killing, are nevertheless divergence-free:
\be
\nabla\cdot {}^{k}X = -{1\over2}G_{AB}\left(
{\cal L}_{{}^{k}X}G\right)^{AB}=
\int d^{3}x\,d^{3}y\,\delta(\bfx-
\bfy)\partial_{x^k}\delta(\bfx-\bfy)=0
\label{eq: spatial translation divergence free}
\ee
since $\partial_{x}\delta(x-y)$ is antisymmetric; similarly for
${}^{k}Y$.  As we shall see later, these results considerably
simplify the Dirac quantization of the Poincar\'{e} algebra.

\par
But first we must work out the algebra at the classical level.  In
terms of our previous notation the Poisson bracket can be expressed
as
\be
\left\{\C_{s}(S),\C_{t}(T)\right\}=\C_{s+t-1}(-[\![S,T]\!]),
\label{eq: classical Poisson algebra}
\ee
where $[\![S,T]\!]$ is the Schouten concomitant \cite{Nijenhuis}
of
$S$ and $T$.  A straightforward, but lengthy
calculation,\footnote{Again, refer to \cite{thesis}.} paying
careful attention to integrations by parts, yields
\bea
\left\{{\cal P}^{\mu},{\cal P}^{\nu}\right\}&=&0
-e\int d^{3}z\,F^{\mu\nu}C_{\gamma},
\label{eq: covariant Poincare algebra 1}
\\
\left\{{\cal J}^{\mu\nu},{\cal P}^{\rho}\right\}&=&
\eta^{\nu\rho}{\cal P}^{\mu}-\eta^{\mu\rho}{\cal
P}^{\nu}-e\int d^{3}z\,\left(z^{\mu}F^{\nu\rho}-
z^{\nu}F^{\mu\rho}\right)C_{\gamma},
\label{eq: covariant Poincare algebra 2}
\\
\left\{{\cal J}^{\mu\nu},{\cal J}^{\rho\sigma}\right\}&=&
\eta^{\mu\sigma}{\cal J}^{\nu\rho}-\eta^{\nu\sigma}{\cal
J}^{\mu\rho}+\eta^{\nu\rho}{\cal J}^{\mu\sigma}-
\eta^{\mu\rho}{\cal J}^{\nu\sigma}
\label{eq: covariant Poincare algebra 3}
\\
&&+e\int d^{3}z \left(
z^{\mu}z^{\sigma}F^{\nu\rho}-z^{\nu}z^{\sigma}F^{\mu\rho}+z^{\nu}
z^{\rho}F^{\mu\sigma}-
z^{\mu}z^{\rho}F^{\nu\sigma}\right)C_{\gamma}.
\nonumber
\eea
Since $C_{\gamma}\approx 0$ defines the constraint surface, we see
that we have explicitly verified the classical Poincar\'{e} algebra
for scalar electrodynamics; we are not aware of any similar
calculation in the literature.
Notice that since
$F^{0j}(\bfz)=-\Pi_{A_{j}(\bfz)}$, the `off-shell pieces', which
are linear combinations of the constraints, are linear and/or
quadratic in the momenta, and these must be
dealt with accordingly when we do Dirac quantization.

\section{Dirac Quantization of Poincar\'{e} Symmetry}

\par
We now proceed with Dirac quantization of the Poincar\'{e} charges,
and subsequent verification of the Poincar\'{e} symmetry at the
quantum level.  The phase space $\Gamma=T^{*}M$, and since $M$
comes equipped with a (flat) metric, $G$, it is natural to choose
Schr\"{o}dinger picture quantization with state space $\F$
consisting of all smooth complex-valued functions on $M$, and
Hilbert space ${\cal H}_{\rm dir}:=L^{2}(M,\bfE)$, the subset of
those which
are square integrable with respect to the volume form $\bfE$
associated with $G$.\footnote{Of course the choice of volume form
is arbitrary---and can be avoided altogether by using
half-densities instead of wavefunctions \cite{Romano}---but this
choice
is made for definiteness.}  The quantization map is not unique:
here we shall choose the simplest one:
\bea
\C_{0}(V)=V&\longmapsto&\Q_{0}(V)=V,
\label{eq: Dirac minimal 0}
\\
\C_{1}(X)=X^{A}P_{A}&\longmapsto&\Q_{1}(X)=-i\hbar\left\{X^{A}
\nabla_{A}+ {1\over2}(\nabla_{A}X^{A})\right\},
\label{eq: Dirac minimal 1}
\\
\C_{2}(K)=K^{AB}P_{A}P_{B}&\longmapsto&\Q_{2}(K)=
(-i\hbar)^{2}\left\{K^{AB}\nabla_{A}\nabla_{B}+
(\nabla_{A}K^{AB})\nabla_{B}\right\}
\nonumber
\\
&&\;\;\;\;\;\;\;\;\;\;=(-i\hbar)^{2}\nabla_{A}K^{AB}\nabla_{B},
\label{eq: Dirac minimal 2}
\\
\C_{3}(T)=T^{ABC}P_{A}P_{B}P_{C}&\longmapsto&\Q_{3}(T)=(-i\hbar)^
{3}\{T^{ABC}\nabla_{A}\nabla_{B}\nabla_{C}
\nonumber
\\
&&+
{3\over2}(\nabla_{A}T^{ABC})\nabla_{B}\nabla_{C}
-{1\over4}(\nabla_{A}\nabla_{B}\nabla_{C}T^{ABC})\},
\label{eq: Dirac minimal 3}
\eea
where, as before, $\nabla$ is the Levi-Civita connection on $M$.
This will be called `minimal quantization' in that given the
leading term (highest order in derivatives) the additional
complimentary terms are the minimum ones necessary to make the
operator self adjoint.  (Notice that cubic operators occur in the
commutator of two quadratic operators, and so are relevant for the
Poincar\'{e} algebra).

\par
In the spirit of Dirac \cite{Dirac} we now account for the
constraints
by quantizing them---on the same footing as any other observable
linear in the momenta: $\hat{C}_{\alpha}=\Q_{1}(\phi_{\alpha})$.
The physical state space, $\Fphys\subset\F$, is then defined as the
collection of states $\Psiphys$ annihilated by the constraint
operators:
\be
\hat{C}_{\alpha}\Psiphys=0\;\forall\alpha\;\Longleftrightarrow\;
\Psiphys\in\Fphys.
\label{eq: quantum constraint}
\ee

\par
As emphasized by Kucha\v{r} \cite{Kuchar1}, the choice of basis for
the gauge vectors $\phi_\alpha$ is arbitrary at the classical
level, but that this breaks down at the quantum level, at least if
one demands that the constraint operators be self adjoint.  The
trouble lies in the complimentary divergence term in
(\ref{eq: Dirac minimal 1}), but can be eliminated by
restricting to a preferred basis which is `compatible' with the
Hilbert space structure:
\be
{\cal L}_{\phi_\alpha}\bfE=0\;\forall\alpha,
\label{eq: fundamental gauge relation}
\ee
i.e. in which the $\phi_\alpha$ are divergence-free.  This
restriction is natural in the sense that
(\ref{eq: quantum constraint}) then implies
$\phi_{\alpha}\Psiphys=0\;\forall\alpha$, i.e. $\Fphys$ consists
of
gauge invariant complex-valued functions on $M$.

\par
Furthermore, given such a basis one is free to transform to any
other basis whose elements are all divergence-free, i.e. taken from
the set
\be
{\cal G}:= \left\{\mu=\mu^{\alpha}\phi_{\alpha}\mid\nabla\cdot\mu=0
\;\Longleftrightarrow\;\phi_{\alpha}\mu^{\alpha}=0\right\}.
\label{eq: gauge vector space}
\ee
In our case the Lagrangian provides a natural basis of
$\phi_\alpha$ which are Killing, and so certainly satisfy
(\ref{eq: fundamental gauge relation}).
Furthermore, constraint operators constructed from elements of
${\cal G}$ will be consistent (first class) iff
$\phi_{\gamma}f^{\gamma}_{\alpha\beta}=0$, or, equivalently,
$f^{\gamma}_{\gamma\beta}=\phi_{\beta}f$ for some scalar $f$.  In
our case the structure functions $f^{\gamma}_{\alpha\beta}$ vanish
(see (\ref{eq: gauge algebra})), and so this condition is trivially
satisfied.

\par
The Poincar\'{e} charges $\Pnot$,$\Pk$,$\Jk$,$\Kk$ contain pieces
zero, first and second order in momenta, and are quantized
accordingly using the minimal quantization scheme.  In order for
the resulting quantum Poincar\'{e} charges to be observables they
must commute with the constraint operators $\Q_{1}(\mu)$ (at least
on $\Fphys$) for all $\mu\in {\cal G}$.

\par
Let the scalar $U$ represent any of the Hamiltonian of boost
potential $V$ or ${}^{k}V$; we have
\be
{1\over{i\hbar}}[\Q_{0}(U),\Q_{1}(\mu)]=
\Q_{0}(-[\![U,\mu ]\!]).
\ee
But $-[\![U,\mu ]\!]={\cal L}_{\mu}U$ vanishes since the potentials
are constant along the gauge orbits.  For a vector $Z$,
representing any of the spatial translation or rotation vectors
${}^{k}X$ or ${}^{k}Y$, we find
\be
{1\over{i\hbar}}[\Q_{1}(Z),\Q_{1}(\mu)]=
\Q_{1}(-[\![Z,\mu]\!]),
\label{eq: 1 observable}
\ee
where $-[\![Z,\mu]\!]={\cal L}_{\mu}Z$.  Using the fact that ${\cal
L}_{\phi_{\alpha}}Z=0$ it is easy to show that ${\cal
L}_{\mu}Z\in{\cal G}$, so the right hand side of
(\ref{eq: 1 observable}) annihilates $\Psiphys$.

\par
Finally, letting $K$ stand for either the inverse metric, $G^{-1}$,
or any of the boost tensors, ${}^{k}K$, we have
\be
{1\over{i\hbar}}[\Q_{2}(K),\Q_{1}(\mu)]=
\Q_{2}(-[\![K,\mu]\!])+\hbar^{2}\Q_{0}(W);
\nonumber
\ee
\be
W={1\over2}\nabla_{A}(K^{AB}\nabla_{B}(\nabla\cdot\mu)).
\label{eq: 21 vanHove}
\ee
The $\hbar^{2}\Q_{0}(W)$ term is a van Hove anomaly (discussed more
fully below), which in this case vanishes precisely because of the
restriction
(\ref{eq: gauge vector space}).  Furthermore,
\be
-[\![K,\mu]\!]={\cal L}_{\mu}K=\mu^{\alpha}{\cal
L}_{\phi_{\alpha}}K-2\psi^{\alpha}\otimes_{\rm S}\phi^{\alpha},
\label{eq: symm}
\ee
where $\otimes_{\rm S}$ denotes symmetrized tensor product.  The
term on the right hand side vanishes, and the Cartesian components
of the vector fields $\psi^{\alpha}$ are $\psi^{\alpha
A}=K^{AB}\nabla_{B}\mu^{\alpha}$.  Quantizing the remaining term
yields an operator proportional to
\be
\nabla_{A}\psi^{\alpha A}\phi_{\alpha}^{B}\nabla_{B}+
\nabla_{A}\phi_{\alpha}^{A}\psi^{\alpha B}\nabla_{B}.
\nonumber
\ee
The first term annihilates $\Psiphys$, and the second is equivalent
to
\be
(\nabla\cdot\phi_{\alpha})\psi^{\alpha}+
\phi_{\alpha}\psi^{\alpha}=
[\phi_{\alpha},\psi^{\alpha}]+
\psi^{\alpha}\phi_{\alpha}.
\ee
Now the second term on the right hand side of this expression
annihilates $\Psiphys$,
and, furthermore, the commutator vanishes:
\be
({\cal L}_{\phi_{\alpha}}\psi^{\alpha})^{A}=
({\cal L}_{\phi_{\alpha}}K)^{AB}\nabla_{B}\mu^{\alpha}+
K^{AB}({\cal L}_{\phi_{\alpha}}\nabla\mu^{\alpha})_{B}=0.
\label{eq: gauge derivative of psi alpha}
\ee
Thus, the quantum Poincar\'{e} charges are observables.

\par
The next question to ask is whether or not they realize the
Poincar\'{e} algebra when acting on physical states.  There are two
considerations:  van Hove anomalies, and whether or not the minimal
quantization of the off-shell pieces in
(\ref{eq: covariant Poincare algebra 1}
--\ref{eq: covariant Poincare algebra 3}) produces operators which
annihilate physical states.  We discuss these in turn.

\par
Since the work of Groenewold \cite{Groenewold} and van Hove
\cite{vanHove} it has been known that no map from classical to
quantum observables exists which preserves the entire Poisson
algebra.\footnote{See, e.g., \cite{Abraham} for a more precise
statement.}  For the minimal quantization map given in
(\ref{eq: Dirac minimal 0}--\ref{eq: Dirac minimal 3}) van Hove
anomalies appear first in the quadratic-linear commutator: refer
to
(\ref{eq: 21 vanHove}), which applies for generic $K$, and $\mu$
replaced by a generic vector field, $Z$.  But the only vector
fields occuring in the Poincar\'{e} charges are the spatial
translation and rotation vectors ${}^{k}X$ and ${}^{k}Y$, which are
both divergence-free
(see \ref{eq: spatial translation divergence free}), and so this
particular van Hove anomaly is not present.

\par
For the generic quadratic-quadratic commutator:
\be
{1\over{i\hbar}}[\Q_{2}(K),\Q_{2}(L)]=\Q_{3}(-[\![K,L]\!])
+\hbar^{2}\Q_{1}(Z),
\label{eq: 2 2 commutator}
\ee
\bea
Z^{D}&:=&{1\over2}\nabla_{B}\nabla_{C}[\![K,L]\!]^{BCD}-
\nabla_{B}A^{BD},
\nonumber
\\
A^{BD}&:=&
K^{AB}L^{CD}\R_{AC}+(\nabla_{C}K^{AB})(\nabla_{A}L^{CD})
\nonumber
\\
&&-{1\over
3}\nabla_{C}(K^{AB}(\nabla_{A}L^{CD})-K^{AD}(\nabla_{A}L^{CB})),
-(K\leftrightarrow L).
\label{eq: W in 2 2 commutator}
\eea
For our example the van Hove term, $\hbar^{2}\Q_{1}(Z)$, vanishes
because the Ricci tensor $\R_{AB}$ is zero ($M$ is flat) and the
inverse metric
and boost tensors are all covariantly constant.
Thus the Poincar\'{e} algebra is free of van Hove anomalies under
minimal Dirac quantization.

\par
We now come to the quantization of the off-shell pieces in
(\ref{eq: covariant Poincare algebra 1}
--\ref{eq: covariant Poincare algebra 3}), of which there are
essentially only two types.  The first type has the form
\be
-e\int d^{3}z\,F^{ij}C_{\gamma}=:\C_{1}(\mu^{\gamma}\phi_{\gamma}),
\ee
where (with $\gamma:=\bfz$) the scalars
\be
\mu^{\gamma}:=-eF^{ij}(\bfz).
\ee
But since the electromagnetic field strength is gauge invariant,
we
certainly have $\phi_{\gamma}\mu^{\gamma}=0$, so
$\mu^{\gamma}\phi_{\gamma}\in {\cal G}$, in which case
$\Q_{1}(\mu^{\gamma}\phi_{\gamma})\Psiphys=0$.

\par
The second type has the form
\be
-e\int d^{3}z\,F^{0j}C_{\gamma}=:
\C_{2}(\psi^{\gamma}\otimes_{\rm S}\phi_{\gamma}),
\ee
where the vector field components
\be
\psi^{\gamma A}=(e\delta^{i}_{j}\delta(\bfz-\bfx),0,0).
\ee
The quantization of this second type of term is exactly analogous
to the discussion following
(\ref{eq: symm}), except the Lie derivative corresponding to
(\ref{eq: gauge derivative of psi alpha}) is
\be
({\cal L}_{\phi_{\gamma}}\psi^{\gamma})^{A}=-\int
d^{3}z\,d^{3}y\,\left\{
e\delta(\bfz-\bfy){{\delta}\over{\delta A_{j}(\bfy)}}\right\}
\phi_{\gamma}^{A}(Q)=0,
\ee
since $\phi_{\gamma}^{A}(Q)$ has no dependence on the field
$A_{j}(\bfy)$ (and $\psi^{\gamma A}$ has no field dependence at
all).

\par
In conclusion, we see that the Poincar\'{e} algebra is,
indeed, realized as quantum operators acting on $\Fphys$ using
the minimal quantization scheme for Dirac quantization.

\section{Reduced Quantization of Poincar\'{e} Symmetry}

\par
Classical reduction is readily achieved by chooing the complete set
of gauge invariant functions
\be
q^{a}=\left(B_{i}(\bfx),\rho(\bfx)\right),
\label{eq: q in scalar electrodynamics}
\ee
where $B_{i}(\bfx):=A_{i}(\bfx)+{1\over
e}\partial_{x^i}\theta(\bfx)$ and $\varphi(\bfx)=:\rho(\bfx)\exp
i\theta(\bfx)$, as coordinates on the reduced configuration space,
$m$.  Since the constraints are linear in the momenta, the reduced
phase space $\gamma=T^{*}m$, with canonical coordinates
$(q^{a},p_{a})$.  An observable $\C_{s}(S)$ on $\Gamma$ maps to
the corresponding physical variable $\Cs_{s}(s):=s^{a_{1}\cdots
a_{s}}(q)p_{a_{1}}\cdots p_{a_{s}}$ on $\gamma$, where the tensor
$s$ on $m$ is the (physical) projection of $S$.

\par
In particular, the projected inverse metric
\be
g^{ab}(q)=
\left(\begin{array}{cc}
     \delta_{ij}-\partial_{x^i}{{1}\over{e^{2}
          \rho^{2}(\bfx)}}\partial_{x^j} & 0 \\
     0 & 1
     \end{array}\right)\delta(\bfx-\bfy).
\label{eq: projected inverse metric}
\ee
The other tensors involved in the Poincar\'{e} charges can
similarly be projected onto $m$, and the resulting reduced
Poincar\'{e} charges on $\gamma$ will obviously realize the
Poincar\'{e} algebra
(cf (\ref{eq: covariant Poincare algebra 1}
--\ref{eq: covariant Poincare algebra 3}) with $C_{\gamma}=0$), a
fact which can be verified by direct calculation.  For the purpose
of our discussion it is sufficient to know only the projected boost
tensors:
\be
{}^{m}k^{ab}(q)=
\left(\begin{array}{cc}
     \delta_{ij}x^{m}-\partial_{x^i}{{x^{m}}\over{e^{2}
          \rho^{2}(\bfx)}}\partial_{x^j} & 0 \\
     0 & x^{m}
     \end{array}\right)\delta(\bfx-\bfy).
\label{eq: projected boost tensors}
\ee
We remark that these are Killing tensors, and are in involution
with each other: $[\![{}^{m}k,{}^{n}k]\!]=0$.  Also it can be shown
that the projected spatial rotation and boost vectors are Killing
with respect to the metric $g$ on $m$, and so necessarily are
(Levi-Civita) divergence-free.

\par
We now quantize the reduced Poincar\'{e} charges, and attempt to
verify the Poincar\'{e} symmetry at the quantum level.  In analogy
with the Dirac quantization considered earlier, we choose
Schr\"{o}dinger picture quantization with Hilbert space ${\cal
H}_{\rm red}:=L^{2}(m,\bfe)$, where $\bfe$ is the volume form
associated with the metric $g$ on $m$.  As was the case with Dirac
quantization, the choice of quantization map is not
unique---especially now that the configuration space is not flat
(see \cite{ii}, and references therein).

\par
But in order to compare Dirac and reduced quantization on an `equal
footing', we again choose minimal quantization
(cf (\ref{eq: Dirac minimal 0}--\ref{eq: Dirac minimal 3})), which
is also in keeping with tradition in the Dirac versus reduced
quantization debate in the literature
\cite{Ashtekar,Kuchar1,Kuchar2,Romano,Schleich,Kunstatter1}.
In particular, for a physical variable quadratic in the momenta:
\be
\Cs_{2}(k)=k^{ab}p_{a}p_{b}\longmapsto\Qs_{2}(k):=
(-i\hbar)^{2}\del_{a}k^{ab}\del_{b}=
(-i\hbar)^{2}\{\partial_{a}k^{ab}\partial_{b}+k^{ab}
(\partial_{a}\ln\omega)\partial_{b}\},
\label{eq: minimal reduced}
\ee
where $\del$ is the Levi-Civita connection on $m$, $\partial_{a}$
is the (functional) derivative with respect to $q^{a}$, and
$\omega=\sqrt{\det g_{ab}}$ is the measure on $m$ in the
coordinates $q^{a}$.

\par
Now, as noted above, the classical Poincar\'{e} charges realize the
Poincar\'{e} algebra, and this will automatically extend to the
quantum level provided there are no van Hove anomalies.  The first
place such an anomaly might arise---with quadratic-linear
commutators (cf
(\ref{eq: 21 vanHove}) with $\nabla\longmapsto\del$, etc.)---it
does not, since the projected translation and rotation vectors are
divergence-free.
This leaves a potential anomaly only with quadratic-quadratic
commutators (cf
(\ref{eq: W in 2 2 commutator})).  However, instead of dealing with
covariant derivatives\footnote{Note that even if $K$ is covariantly
constant on $M$, its physical projection $k$ on $m$ need not be.}
and the Ricci tensor on $m$ we calculate, in terms of $\omega$:
\bea
{1\over{(-i\hbar)^4}}[\Qs_{2}(k),\Qs_{2}(l)]&=&[\![k,l]\!]^{bcd}
\partial_{b}\partial_{c}\partial_{d}+{3\over2}\left\{(\partial_{b}
[\![k,l]\!]^{bcd})+
(\partial_{b}\ln\omega)[\![k,l]\!]^{bcd}\right\}
\partial_{c}\partial_{d}
\nonumber
\\
&&+\left\{k^{ab}\partial_{a}(\partial_{b}(v^{d}))+
u^{b}\partial_{b}(v^{d})-(k\leftrightarrow l)\right\}\partial_{d},
\label{eq: quadratic quadratic explicitly on reduced}
\eea
for generic $k$ and $l$, where the `vectors'
\bea
u^{b}&:=&\partial_{a}k^{ab}+(\partial_{a}\ln\omega)k^{ab},
\\
v^{d}&:=&\partial_{c}l^{cd}+(\partial_{c}\ln\omega)l^{cd}.
\eea
If $k$ and $l$ represent any of either the inverse metric or boost
tensors then $[\![k,l]\!]^{bcd}=0$, as noted earlier---a fact which
must be true for the Poincar\'{e} algebra to close.  Thus, for the
quantum commutator to vanish, as it should, we require all
components of the terms in braces in the last line of
(\ref{eq: quadratic quadratic explicitly on reduced}), which we
will denote as $\zeta^{d}$, to be zero.

\par
For instance, the (quadratic-quadratic part of) the boost-boost
commutator corresponds to taking $k={}^{m}k$, $l={}^{n}k$, and we
find, using
(\ref{eq: projected boost tensors}), that only the $d=\rho(\bfw)$
component of $\zeta$ is potentially nonvanishing:
\bea
\zeta^{\rho(\bfw)}&=&\int
d^{3}x\,d^{3}y\,y^{m}w^{n}\delta(\bfx-\bfy)
{{\delta^{3}\ln\omega}\over
{\delta\rho(\bfx)\delta\rho(\bfy)\delta\rho(\bfw)}}
\nonumber
\\
&&+\int d^{3}y\,y^{m}w^{n}
{{\delta\ln\omega}\over{\delta\rho(\bfy)}}
{{\delta^{2}\ln\omega}\over
{\delta\rho(\bfy)\delta\rho(\bfw)}}-(m\leftrightarrow n)
\label{eq: zeta}
\eea
(in uncondensed notation).

\par
In order to evaluate $\delta\ln\omega$ we first observe that (see
(\ref{eq: projected inverse metric}))
\be
\det g^{ab}={\rm Det}\left[\delta_{ij}-\partial_{i}
{{1}\over{e^{2}\rho^{2}}}\partial_{j}\right]
={\rm Det}\left[{{1}\over{\rho^{2}}}
\left(-{{1}\over{e^{2}}}\partial^{2}+\rho^{2}\right)\right],
\ee
where Det denotes functional determinant, and
$\partial^{2}:=\partial_{i}\partial_{i}$.  The last equality
follows by decomposing the eigenvectors of the operator
$\delta_{ij}-...$ into its transverse and longitudinal parts, and
examining the eigenvalues.  Hence
\bea
\delta\ln\omega&=&-{1\over2}\delta\ln {\rm
Det}\left[{{1}\over{\rho^{2}}}
\left(-{{1}\over{e^{2}}}\partial^{2}+\rho^{2}\right)\right]=
-{1\over2}\delta {\rm Tr}\ln\left[{{1}\over{\rho^{2}}}
\left(-{{1}\over{e^{2}}}\partial^{2}+\rho^{2}\right)\right]
\nonumber
\\
&&=-{1\over2}\delta {\rm Tr}\ln{{1}\over{\rho^{2}}}
-{1\over2}\delta {\rm Tr}\ln
\left(-{{1}\over{e^{2}}}\partial^{2}+\rho^{2}\right)
=:\delta I +\delta I\! I,
\label{eq: delta ln omega split}
\eea
where we have assumed that the functional trace, Tr, satisfies the
usual
cyclicity property, and that $\delta {\rm Tr}\ln A={\rm Tr}A^{-1}
\delta A$ for any operator $A$.

\par
The first term is straightforward to evaluate:\footnote{We are
being somewhat cavalier about regularization---simply because it
is
difficult to do much better---but we believe that our final
conclusions still carry sufficient weight to be of interest, as we
shall argue.}
\be
\delta I=\int d^{3}x\,\delta\rho(\bfx)\left\{
{{\delta( \ZERO )}\over{\rho(\bfx)}}\right\},
\label{eq: delta I}
\ee
but the second term is more difficult.  Following Hawking's
discussion \cite{Hawking} on zeta function regularization we write
\be
\delta I\! I={{d}\over{ds}}\left.\left[
{{1}\over{\Gamma(s)}}\int d^{3}x\,\int_{0}^{\infty}d\tau\,
\tau^{s}\delta D\,K(\bfx,\bfx,\tau)\right]\right|_{s=0}.
\label{eq: delta IIa}
\ee
Here the positive definite operator
\be
D:=-{{1}\over{e^{2}}}\partial^{2}+\rho^{2}+\epsilon,
\ee
where $\epsilon>0$ is a regulating `mass' parameter. Its
associated heat kernel, $K(\bfx,\bfy,\tau)$, satisfies
\be
{{\partial}\over{\partial\tau}}K(\bfx,\bfy,\tau)
+DK(\bfx,\bfy,\tau)=0,
\label{eq: heat}
\ee
with initial condition $K(\bfx,\bfy,0)=\delta(\bfx-\bfy)$.  $D$
(and $\delta D$ in
(\ref{eq: delta IIa})) act on the first argument of $K$.  In our
case $\delta D$ is simply $2\rho\delta\rho$.

\par
As in, e.g., \cite{DeWitt} we factorize:
\be
K(\bfx,\bfy,\tau)=:K_{0}(\bfx,\bfy,\tau)\Lambda(\bfx,\bfy,\tau)
\ee
into a singular piece
\be
K_{0}(\bfx,\bfy,\tau)=\left({{e}\over{4\pi\tau}}\right)^{3\over2}
\exp\left(-{{e^{2}|\bfx-\bfy|^{2}}\over{4\tau}}-
\epsilon\tau\right),
\ee
which satisfies
(\ref{eq: heat}) with $\rho^{2}\equiv 0$, and initial condition
$K_{0}(\bfx,\bfy,0)=\delta(\bfx-\bfy)$, and a regular piece,
$\Lambda(\bfx,\bfy,\tau)$.  The latter contains the $\rho$
dependence of $K$, and satisfies
\be
{{\partial}\over{\partial\tau}}\Lambda(\bfx,\bfy,\tau)
+{{1}\over{\tau}}(x^{i}-
y^{i})\partial_{x^{i}}\Lambda(\bfx,\bfy,\tau)
=-\left(-{{1}\over{e^{2}}}\partial_{x}^{2}+
\rho^{2}(\bfx)\right)\Lambda(\bfx,\bfy,\tau),
\ee
with initial condition $\Lambda(\bfx,\bfx,0)=1$.  Again as in
\cite{DeWitt}, we now expand
\be
\Lambda(\bfx,\bfy,\tau)=\sum_{n=0}^{\infty}a_{n}(\bfx,\bfy)
\tau^{n},
\ee
and find that the coefficients $a_n$ satisfy the recursion relation
\be
na_{n}+(x^{i}-y^{i})\partial_{x^{i}}a_{n}+\left(-
{{1}\over{e^{2}}}\partial_{x}^{2}+
\rho^{2}(\bfx)\right)a_{n-1}=0,\;n=1,2,\ldots
\label{eq: recursion}
\ee
with $a_{0}(\bfx,\bfy)=1$.  In the coincidence limit we obtain
\bea
a_{1}(\bfx,\bfx)&=&-\rho^{2}(\bfx),
\\
a_{2}(\bfx,\bfx)&=&{1\over2}\left(-
{{1}\over{3e^{2}}}\partial_{x}^{2}+
\rho^{2}(\bfx)\right)\rho^{2}(\bfx),
\eea
in agreement with \cite{Nepomechie}.

\par
Substituting these results into
(\ref{eq: delta IIa}) yields an expression valid for ${\rm
Re}(s)>\frac{1}{2}$, which when analytically continued to $s=0$
gives
\be
\delta I\! I=\int d^{3}x\,\delta\rho(\bfx)\left\{
-\sum_{n=0}^{\infty}c_{n}\rho(\bfx)a_{n}(\bfx,\bfx)\right\},
\label{eq: delta IIb}
\ee
where the coefficients
\be
c_{n}=\left({{e}\over{4\pi}}\right)^{3\over2}
{{\Gamma(n-\frac{1}{2})}\over{\epsilon^{n-\frac{1}{2}}}}.
\ee
Now although $a_{n}(\bfx,\bfx)$ effectively goes like $1/n!$ (see
(\ref{eq: recursion})), the series in
(\ref{eq: delta IIb}) nevertheless diverges as
$\epsilon\longrightarrow 0$, and must therefore be treated as a
formal expansion in powers of $\epsilon$.  (And note also that
powers of $e$ are associated with derivatives of $\rho$).

\par
Finally, working out the higher order variations of $I$ and $I\!
I$, and using these results in
(\ref{eq: zeta}), yields
\be
\zeta^{\rho(\bfw)}={{2c_{2}}\over{3e^{2}}}w^{n}\left\{
{{\delta\ln\omega}\over{\delta\rho}}\rho(\partial_{m}\rho)+
\left(\partial_{m}{{\delta\ln\omega}\over{\delta\rho}}\right)
\rho^{2}\right\}-(m\leftrightarrow n)
\ee
(plus higher order terms).  It is instructive to note that if
$\delta\ln\omega/\delta\rho\propto\rho^{k}$, the term in braces
vanishes iff $k=-1$, a situation which corresponds precisely to
term $I$ of $\ln\omega$ (see
(\ref{eq: delta I})).  Term $I\! I$, on the other hand, contributes
a polynomial with higher powers of $\rho$ (and also derivatives of
$\rho$):  the leading order (in both $\epsilon$ and $e$)
contribution is
\be
\zeta^{\rho(\bfw)}={{1}\over{(12\pi)^{2}}}\,{{e}\over{\epsilon}}\,
(w^{n}\partial_{w^m}-w^{m}\partial_{w^n})\rho^{3}(\bfw),
\ee
which does not vanish for generic $\rho$.
Also note that higher order terms in the expansion
(\ref{eq: delta IIb}) do not contain this combination of $e$ and
$\epsilon$, so no cancelation of this piece is possible.

\par
Thus, the boost-boost commutator (and hence the Poincar\'{e}
algebra as a whole) fails to be realized at the quantum level using
minimal reduced quantization.

\section{Discussion}

\par
We have thus shown that minimal Dirac quantization preserves the
Poincar\'{e} symmetry of scalar electrodynamics, but that minimal
reduced
apparantly does not.  To better understand how this comes about it
is instructive to determine what the Dirac-quantized Poincar\'{e}
charges
look like acting on physical states, so they can be compared on the
same footing with their reduced counterparts.

\par
For instance, direct calculation using
(\ref{eq: metric}), (\ref{eq: Dirac minimal 2}), and
(\ref{eq: q in scalar electrodynamics}) shows that the kinetic
energy operator $\Q_{2}(\frac{1}{2}G^{-1})$, acting on
$\Psiphys\in\Fphys$, is equivalent to $\Qs_{2}(\frac{1}{2}g^{-1})$
(see (\ref{eq: minimal reduced})), except with
$\partial_{a}\ln\omega$ replaced by an object we call, similarly,
$\partial_{a}\ln\omega^{'}$, whose only nonvanishing component is
\be
{{\delta\ln\omega^{'}}\over{\delta\rho(\bfx)}}=
{{\delta( \ZERO )}\over{\rho(\bfx)}}.
\label{eq: delta ln omega prime}
\ee
In fact, the analogous statement applies for the entire set of
Poincar\'{e} charges:  minimal Dirac quantization (acting in
$\Fphys$) is identical in {\em form} with minimal reduced
quantization, except with $\partial_{a}\ln\omega^{'}$ in place of
$\partial_{a}\ln\omega$, a difference which corresponds to
retaining only the first term, $\delta I$, in
(\ref{eq: delta ln omega split}).  (Compare
(\ref{eq: delta ln omega prime}) with (\ref{eq: delta I})).

\par
This means, for instance, that the quadratic-quadratic commutator
in minimal Dirac quantization has the same form as
(\ref{eq: quadratic quadratic explicitly on reduced}), but when
applied to the boost-boost commutator is easily seen to yield
$\zeta=0$, i.e. no anomaly, as expected from the results of
section~4.  We remark that, although the term $\delta I$ in
(\ref{eq: delta I}) contains $\delta ( \ZERO )$, and so is not
regulated, it is {\em common} to both the Dirac and reduced
approaches, and the (independent) results of section~4 support the
proposition that this term does not cause a problem with the
Poincar\'{e} algebra.  Rather, it is the additional term, $\delta
I\! I$, present in reduced quantization---in particular, those
pieces involving {\em derivatives} of $\rho$, which
begin to appear with the $n=2$ term in
(\ref{eq: delta IIb})---that causes a van Hove anomaly.

\par
In fact, we observe that $\exp(-I\! I)$ is nothing but the volume
element, $\sqrt{\det\gamma_{\alpha\beta}}$, on the gauge orbits,
where the metric
\be
\gamma_{\alpha\beta}:=G_{AB}\phi_{\alpha}^{A}\phi_{\beta}^{B}
=\left(-
{{1}\over{e^{2}}}\partial_{x}^{2}+
\rho^{2}(\bfx)\right)\delta(\bfx-\bfy).
\label{eq: metric on orbits}
\ee
Now in must be emphasized that, in general, the minimal Dirac and
minimal reduced quantization schemes are {\em not}
equivalent\footnote{For example, the respective Hamiltonians have
different spectra, in general.}
\cite{Ashtekar,Kuchar1,Kuchar2,Romano,Schleich,Kunstatter1}:
Having transformed to a common Hilbert space, the inequivalence
manifests itself in the quadratic operators as a factor ordering
ambiguity involving precisely the above volume element on the gauge
orbits \cite{Kunstatter1}.  Furthermore, for a given model it can
happen that both the Dirac and reduced factor orderings are self
consistent---the relevant example here being Kucha\v{r}'s helix
model \cite{Kuchar2} (which is a finite dimensional analogue of
scalar electrodynamics).
So even though the Hamiltonians might have different spectra, which
could, in principle, be measured, there may be no {\em internal}
physical criterion with which to select the correct factor
ordering, as happens in the helix model \cite{Kuchar2}.
\par
The significant point here is that scalar electrodynamics has an
additional symmetry---the Poincar\'{e} symmetry---and, at the
quantum level, this symmetry is {\em sensitive} to this difference
in factor ordering (or presence of
$\sqrt{\det\gamma_{\alpha\beta}}$), suggesting, in fact, that
minimal Dirac quantization in correct, and minimal reduced is not
(at least in this case).
\par
This result also supports previous work \cite{i,ii} suggesting a
preference for minimal Dirac over minimal reduced
because of the natural similarity of the former with several
curved-space quantization schemes proposed in the literature.
\par
In general, then, demanding the preservation of a sufficiently
nontrivial classical symmetry at the quantum level may serve as a
useful internal physical criterion with which to select amongst
inequivalent factor orderings, as we have demonstrated here. It
might also be instructive to find a suitable finite-dimensional
model with which to demonstrate this point, free of any
regularization complications.

\begin{center}
\vspace{10 pt}
{\large\bf Acknowledgements}
\vspace{0 pt}
\end{center}

\par
G.K. would like to thank D.J. Toms for useful discussions during
the early stages of this work. R.J.E. would like to express his
gratitude to S. Carlip and A.O. Barvinsky for helpful discussions
concerning regularization, as well as acknowledge the financial
support of a University of  Manitoba Graduate Fellowship. This work
was supported in part by
the Natural Sciences and Engineering Research Council of Canada.

\end{document}